\documentclass[aps,prc,onecolumn,superscriptaddress]{revtex4}
\usepackage[latin1]{inputenc}
\usepackage{bm}
\usepackage{epsfig}
\usepackage{amsthm}
\usepackage{amsfonts}
\usepackage{float}
\usepackage{amsmath,amssymb}
\usepackage{color}
\usepackage{slashed}
\usepackage{relsize}
\usepackage[toc,page]{appendix} 
\usepackage{wrapfig}
\usepackage{graphicx}
\usepackage{dcolumn}
\usepackage{bm}
\newcommand{\be}{\begin{equation}} 
\newcommand{\ee}{\end{equation}} 
\newcommand{\bea}{\begin{eqnarray}} 
\newcommand{\eea}{\end{eqnarray}} 
\newcommand{\nn}{\nonumber} 
\newcommand{\mintedim}[2]{{\int\kern-0.50em\mbox{{\small$\mathop{\frac{\mbox{{\small${\rm d^{#2}}\vect{#1}$}}}{\mbox{{\small$(2\pi)^{#2}$}}}}$}}\ }} 
\newcommand{\inteonedim}[1]{{\int_0^\infty\kern-1em\mbox{{\small${\rm d}{#1}$}}}} 
\newcommand{\vect}[1]{\bm{#1}} 

\begin{document}
\title{
\textcolor{black} {Rapidity spectra in high-energy collisions and longitudinal nuclear suppression from nonadditive statistics}
}

\author{Trambak Bhattacharyya}
\email{trambak.bhattacharyya@ujk.edu.pl}
\affiliation{Institute of Physics, Jan Kochanowski University, Kielce 25-406, Poland}

\author{Maciej Rybczy\'{n}ski}
\email{maciej.rybczynski@ujk.edu.pl}
\affiliation{Institute of Physics, Jan Kochanowski University, Kielce 25-406, Poland}

\author{Zbigniew W{\l}odarczyk}
\email{zbigniew.wlodarczyk@ujk.edu.pl}
\affiliation{Institute of Physics, Jan Kochanowski University, Kielce 25-406, Poland}

\begin{abstract} 
\vspace{10pt}
We investigate the longitudinal nuclear suppression factor defined by a scaled ratio of rapidity distributions. To study this experimental observable, we describe three approaches involving numerical and analytical calculations. We first approach this problem by conducting model studies using EPOS, FTFP$_{\text{BERT}}$, and HIJING, and notice that while EPOS shows a decreasing trend of \textcolor {black} {longitudinal suppression} at forward/backward rapidities, the latter two models display an \textcolor {black} {increase} of the ratio. The analytical approaches involve, first, the quasi-exponential distribution obtained from the Tsallis statistics, and second, the nonadditive Boltzmann transport equation in the relaxation time approximation. We notice that our analytical results satisfactorily describe NA61 experimental data (for $\sqrt{s_{NN}}$=6.3, 7.6, 8.8, 12.3, and 17.3 GeV) for the negatively charged pions.
\end{abstract}

\maketitle
\section{Introduction and findings from numerical models}
Particle spectra are important tools to study the dynamics of high-energy collisions. It has been shown \textcolor {black} {in many studies} that transverse momentum distributions and rapidity spectra at various collision energies follow the $q$-exponential and the $q$-Gaussian distribution \cite{MRZWepjc,Cleymansdndy}. Such distributions appear owing to the 
constrained (the first or second moment constraints, \textcolor {black} {i.e., fixed average energy or fixed average squared energy}) maximization of generalized entropy proposed by C. Tsallis \cite{Tsal88}. Such an entropy describes systems having power-law stationary states due to fluctuation, long-range correlation, and anomalous diffusion
\textcolor{black} {\cite{Wilk00,wilkosada,megiasplb,TBplb}.}

Not only spectra, but also their ratios are important tools of studying the dynamics of high-energy collisions. For example, the nuclear suppression factor, defined as the scaled ratio of a spectrum from a heavy-ion collision to that from a proton-proton collision, helps us determine if the heavy-ion collisions are mere superpositions of the 
proton-proton collisions \textcolor{black} {\cite{raadef}}. Such a ratio provides information about nuclear stopping power in the transverse plane. 

In the longitudinal direction, nuclear stopping implies a shift of the rapidity distribution towards  mid-rapidity. However, such a distribution has a strong collision energy dependence. For example, at the AGS energies there is a peak in the net-proton rapidity spectrum \cite{E9172001,E8021999,E8772000}, at SPS energy there is a dip \cite{NA491999}, and at RHIC, it is almost flat with small peaks near the beam rapidity \cite{BRAHMS2004}. Such behaviour indicates that with higher energy, incident nuclei do not lose their energy, but pass through the target. 

In this article, we study a suppression/modification factor in the longitudinal plane that is experimentally obtained from the following ratio:
\bea
R_{dN/dy} = \frac{\frac{dN^{\text{Pb-Pb}}}{dy}}{\left\langle N_{\text{coll}}\right\rangle \frac{dN^{pp}}{dy}},
\eea
where $dN/dy$ is the rapidity ($y$) distribution and $\langle N_{\text{coll}} \rangle$ is the average number of nucleon-nucleon binary collisions when two heavy-ions (in this case, Pb) collide. 
\textcolor{black} {This observable is different from the commonly discussed $R_{\text{AA}}$, and provides information about longitudinal physics. 
Starting from the joint distribution $d^2N/dp_{\text{T}}dy$, the quantity $R_{dN/dy}$ is defined by the marginal distributions $dN/dy$, while $R_{\text{AA}}$ is defined 
by the marginal distributions $dN/dp_{\text{T}}$. These marginals examine different aspects of underlying dynamics because each integrates out a different variable from the parent joint distribution.
}

Qualitatively, we expect that $R_{dN/dy}$ will show a dip near mid-rapidity and will increase at higher forward/backward rapidity due to a low density of the QGP medium as $y$ increases. To verify this argument, \textcolor{black} {and to investigate the behaviour and sensitivity of $R_{dN/dy}$}, we conducted numerical \textcolor{black} {calculations} with the help of three models, namely, EPOS, FTFP$_{\text{BERT}}$, and HIJING for Pb-Pb and p+p collisions at $\sqrt{s_{NN}}$=7.6 GeV, and 17.3 GeV. \textcolor{black} {The} EPOS model follows a consistent quantum mechanical multiple scattering approach based on partons and strings, where cross sections and the particle production are calculated considering energy conservation. \textcolor{black} {The resulting initial conditions are evolved utilizing a hydrodynamic description of the medium} \cite{Werner:2005jf}. FTFP$_{\text{BERT}}$ is a hadronic physics model in Geant4, providing a comprehensive simulation of hadronic showers in collider and detector physics, that combines the Fritiof (FTF) string model for high-energy interactions ($>$ 4-5 GeV) with the Bertini-style intra-nuclear cascade (BERT) model for low-energy interactions ($<$ 5 GeV) \cite{ref:geant4-website}. HIJING is a Monte Carlo event generator that incorporates both perturbative QCD processes and soft interactions \cite{HIJING}.

\begin{figure}[!htb]
\minipage{0.48\textwidth}
\begin{center}
\hspace{-0.in}
\includegraphics[width=\linewidth]{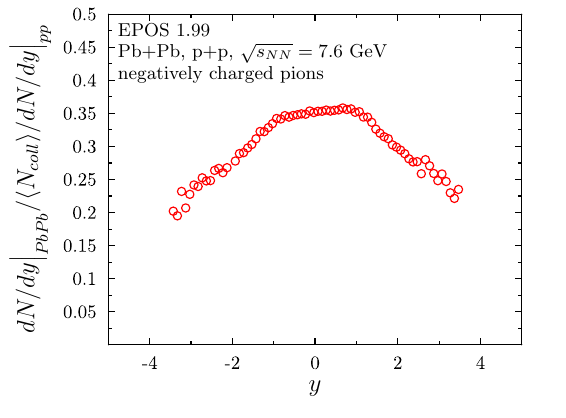}
\caption{Nuclear suppression factor of the $\pi^{-}$ at $\sqrt{s}$=7.6 GeV calculated from EPOS.}
\label{rdndyepos7pt6}
\end{center}
\endminipage\hfill
\minipage{0.48\textwidth}
\vspace*{0.1cm}
\hspace*{2pt}
\includegraphics[width=\linewidth]{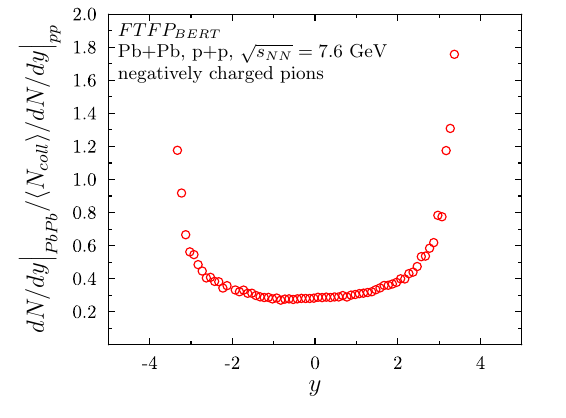}
\vspace{-12pt}
\caption{Nuclear suppression factor of the $\pi^{-}$ at $\sqrt{s}$=7.6 GeV calculated from FTFP$_{\text{BERT}}$.}
\label{rdndyftfp7pt6}
\endminipage\hfill
\end{figure}

\begin{figure}[!htb]
\minipage{0.48\textwidth}
\begin{center}
\hspace{-0.in}
\includegraphics[width=\linewidth]{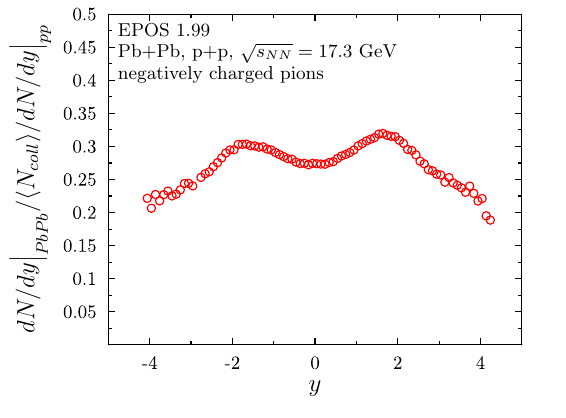}
\caption{The longitudinal nuclear suppression factor of the $\pi^{-}$ at $\sqrt{s}$=17.3 GeV calculated from EPOS.}
\label{rdndyepos17pt3}
\end{center}
\endminipage\hfill
\minipage{0.48\textwidth}
\vspace*{0.2cm}
\hspace*{-0cm}
\includegraphics[width=\linewidth]{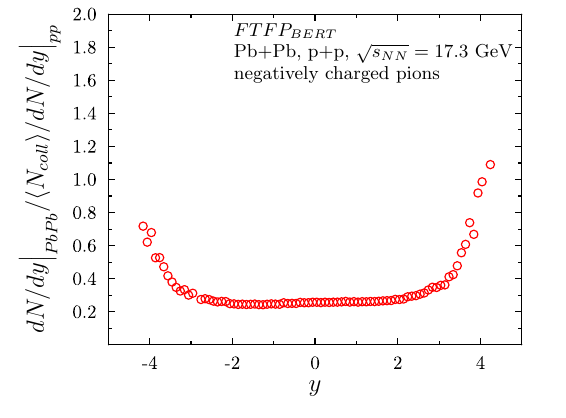}
\hspace{-0cm}
\caption{The longitudinal nuclear suppression factor of the $\pi^{-}$ at $\sqrt{s}$=17.3 GeV calculated from FTFP$_{\text{BERT}}$.}
\label{rdndyftfp17pt3}
\endminipage\hfill
\end{figure}

\begin{figure}[htbp]
\begin{center}
\includegraphics[width=0.45\textwidth]{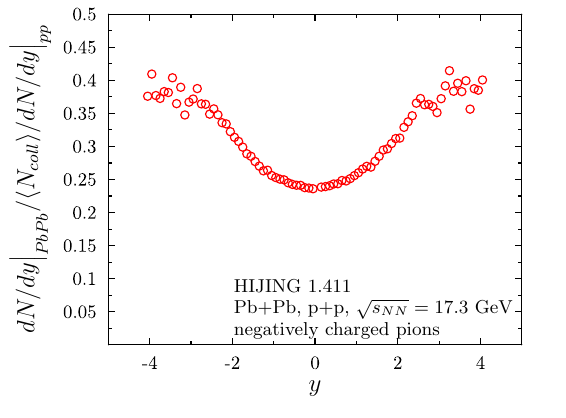}
\caption{The longitudinal nuclear suppression factor of the $\pi^{-}$ at $\sqrt{s}$=17.3 GeV calculated from HIJING.}.
\label{rdndyhijing17pt3}
\end{center}
\end{figure}

The results of simulation are shown in Figs. \ref{rdndyepos7pt6}-\ref{rdndyhijing17pt3}. \textcolor{black} {If we compare Figs. \ref{rdndyftfp17pt3} and \ref{rdndyhijing17pt3} with Fig. \ref{rdndyepos17pt3}, we observe that while the results from FTFP$_{\text{BERT}}$ and HIJING follow our expectation, those from EPOS display a different behaviour. Even at 7.6 GeV, EPOS differs from FTFP$_{\text{BERT}}$.
This difference is also observed when we compare experimental data (Fig. \ref{rdndybte17pt3}) with EPOS calculations. We are not aware of any other work that has reported this difference. Since the EPOS model is widely used by researchers, the reason behind this difference shown by the EPOS model needs to be investigated. However, we reserve this work for the future. 
} 

\textcolor{black}
{The discrepancy among the simulations motivates us to explore analytical models, inspired by nonadditive statistics, that can describe experimental data for meaningful values of parameters.}
This work proposes two different analytical models. The first model involves the phenomenological nonadditive quasi-exponential distribution that describes particle transverse momentum spectra. By integrating out the transverse part, one obtains the rapidity spectra. The other approach involves transport of particles inside a medium. We consider a generalized Boltzmann transport equation that provides a quasi-exponential stationary state and solve it using the relaxation time approximation based on the method proposed in Ref.~\cite{TBPhysica}. \textcolor{black}
{We observe that the analytical models we propose describe experimental data for meaningful values of physically interpretable parameters. 
In summary, the logical flow of the paper is as follows: a) First we carry out simulations as they are important to identify geometric baseline for our observable. b) However, we observe an inconsistency among model results. c) Hence we explore analytical modelling and observe that experimental data can be described using meaningful values of physically interpretable parameters.}
The rest of the paper will be devoted to description of analytical models, analyzing experimental data with those models, and discussions.

\section{$dN/dy$ Ratio from a single-particle distribution}
 \textcolor{black} {To calculate rapidity spectra, we integrate transverse momentum spectra over the transverse momentum variable $p_{\text{T}}$. It has been observed in the literature \cite{MRZWepjc}} that transverse momentum spectra in high-energy collisions are described using the following quasi-exponential single-particle distribution obtained from nonadditive (NA) statistics:
\bea
f_{\text{sp}} = \left(1+(q-1)\frac{E-\mu}{T} \right)^{-\frac{q}{q-1}},
\label{fsp}
\eea
where $q$ is the entropic parameter, $T$ is temperature, $\mu$ is chemical potential, and $E=\sqrt{p^2+m^2}$ is the single-particle energy of a particle of the mass $m$ and 3-momentum $p\equiv |\vec{p}|$. 

\textcolor{black} 
{Following Ref.~\cite{cleymansplb}, we can express the invariant yield $Ed^3N/d^3p$ in terms of the single-particle distribution $f_{\text{sp}}$. Now we separate the 3-momentum volume in transverse and longitudinal part, and express energy and longitudinal momentum ($p_z$) in terms of transverse mass $m_{\text{T}}=\sqrt{p_{\text{T}}^2+m^2}$ and rapidity $y$ as:
$E=m_{\text{T}} \cosh y;~p_z=m_{\text{T}} \sinh y$.} Then, utilizing the single-particle distribution in Eq.~\eqref{fsp}, transverse momentum spectra can be expressed as:
\bea
&&\frac{\textcolor{black}{d^2N}}{dp_{\text{T}}dy} =\frac{gV}{(2\pi)^2} p_{\text{T}} ~m_{\text{T}}\cosh y ~f_{\text{sp}} \nn\\
&&\Rightarrow \frac{dN}{dy} = \int \frac{gV}{(2\pi)^2} p_{\text{T}} ~m_{\text{T}}\cosh y ~f_{\text{sp}} ~dp_{\text{T}}.
\label{dndypheno}
\eea
Putting Eq.~\eqref{fsp} in Eq.~\eqref{dndypheno} we obtain (setting $\mu=0$),
\bea
\frac{dN}{dy} = \int \frac{gV}{(2\pi)^2} p_{\text{T}} ~m_{\text{T}}\cosh y ~\left(1+(q-1)\frac{m_{\text{T}} \cosh(y)}{T} \right)^{-\frac{q}{q-1}}~dp_{\text{T}}.
\label{dndypheno1}
\eea
Now, we analytically calculate a closed form of Eq.~\eqref{dndypheno1} using the Mellin-Barnes contour integral representation \cite{Tsmbprd} of the \textcolor{black} {power-law function 
$\left[1+(q-1)\frac{m_{\text{T}} \cosh(y)}{T} \right]^{-\frac{q}{q-1}}$. Considering  $Y=1$ and
\bea
X 
&=& 
(q-1) \frac{m_{\text{T}}\cosh(y)}{T} \nn\\
&=& 
m(q-1) \frac{\cosh(y) \sqrt{1+k^2}}{T}~~~\left(k=\frac{p_{\text{T}}}{m}\right),
\eea
the Mellin-Barnes contour integral representation of the power-law function can be written in the following form:
\bea    
\frac{1}{(X+Y)^\lambda}=\frac{1}{2\pi i}\int_{\epsilon-i\infty}^{\epsilon+i\infty}\frac{\Gamma(-z)\Gamma(z+\lambda)}{\Gamma(\lambda)}\frac{Y^z}{X^{\lambda+z}}\ dz,
\label{mbrep}
\eea
where $\mathrm{Re}(\lambda)>0$ \& $\mathrm{Re}(\epsilon)\in (-\mathrm{Re}(\lambda),0)$, and the condition is satisfied because $\lambda= q/(q-1)>0$ $\Leftrightarrow$ $q>1$. 
Using Eq.~\eqref{mbrep}, Eq.~\eqref{dndypheno1} is transformed into a product of two independent integrals involving the complex variable $z$ and the scaled variable $k$}:
\bea
\frac{dN}{dy} = 
\frac{gV \cosh(y) ~m^3}{8\pi^3i}
\left( \frac{T \text{sech}(y)}{m(q-1)} \right)^{\frac{q}{q-1}}
\int_{\epsilon-i\infty}^{\epsilon+i\infty} dz \frac{\Gamma(-z)\ \Gamma\left(z+\frac{q}{q-1}\right)}{\Gamma\left(\frac{q}{q-1}\right)}
\left( \frac{T \text{sech}(y)}{m(q-1)} \right)^{z}
\int_{0}^{\infty} k (k^2+1)^{\frac{1}{2}-\frac{z}{2}-\frac{q}{2(q-1)}} dk.
\eea
Performing the integration on $k$, and wrapping the contour in the counter-clockwise direction (so that the poles of
$\Gamma(-z)$ contribute to the contour integration), we obtain the following result:
\bea
\frac{dN}{dy} = 
\frac{g m^2 T V}{4 \pi ^2
   (3-2 q)}
   \left(\frac{T
   \text{sech}(y)}{m
   (q-1)}\right)^{\frac{1}{q-1}}
    \,
   _2F_1\left(\frac{3-2
   q}{q-1},\frac{q}{q-1};\frac{2-q}{q-1};\frac{T
   \text{sech}(y)}{m-m q}\right),
   \label{dndyana}
\eea
where $_2F_1$ is the Hypergeometric function \cite{abst}. 

According to Eq.~\eqref{dndyana}, the $dN/dy$ ratio can be written as:
\bea
R^{(\text{ph})}_{dN/dy} = 
\frac{T_2 V_2 (3-2q_1)}{N_{\text{coll}}T_1 V_1 (3-2q_2)}
\left\{
\frac{T_2}
{(q_2-1)}
\right\}
^{\frac{1}{q_2-1}}
\left\{
\frac{T_1}
{(q_1-1)}
\right\}
^{\frac{1}{1-q_1}}
\frac{\,
   _2F_1\left(\frac{3-2
   q_2}{q_2-1},\frac{q_2}{q_2-1};\frac{2-q_2}{q_2-1};\frac{T_2
   \text{sech}(y)}{m-m q_2}\right)}
   {\,
   _2F_1\left(\frac{3-2
   q_1}{q_1-1},\frac{q_1}{q_1-1};\frac{2-q_1}{q_1-1};\frac{T_1
   \text{sech}(y)}{m-m q_1}\right)}.
   \label{rdndyana}
\eea

\begin{figure}[htbp]
\begin{center}
\includegraphics[width=0.45\textwidth]{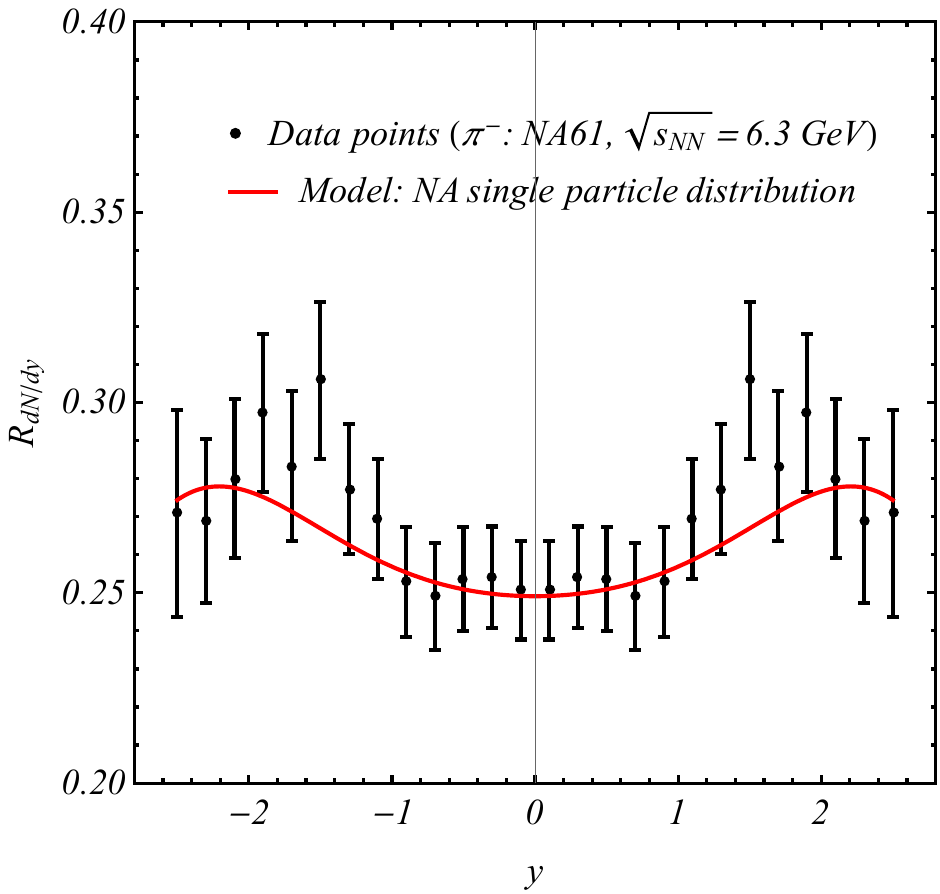}
\caption{Comparison of the ratio of $dN/dy$ in Pb-Pb and p-p collisions ($\sqrt{s}=6.3$ GeV) for the $\pi^{-}$ particles ($m$=0.139 GeV) with Eq.~\eqref{rdndyana} for the following parameter values: $q_1$=1.12, $q_2$=1.10, $T_1$=0.0863 GeV, $T_2$= 0.101 GeV, $V_1$= 75.9 GeV$^{-3}$, $V_2$= 10$^4$ GeV$^{-3}$. We have put $N_{\text{coll}}$=808 calculated from GLISSANDO \cite{glissando}.}
\label{rdndypheno6pt3}
\end{center}
\end{figure}

In Fig.~\ref{rdndypheno6pt3}, we compare Eq.~\eqref{rdndyana} with experimental values of the ratio obtained from experimental data \cite{NA61pp2017,NA49PbPb2}. Although Eq.~\eqref{rdndyana} reproduces the increasing
trend of the ratio at higher rapidity, the model does not explicitly involve evolution of the initial distribution due to interaction. As a starting point, we may study such an evolution using 
the relaxation time approximation of nonadditive Boltzmann transport equation (NABTE).

\section{$dN/dy$ ratio from Nonadditive Boltzmann transport equation}
\subsection{NABTE in the relaxation time approximation}
If at time $t=0$, all the external forces are switched off and the gradient is cancelled, the nonadditive Boltzmann 
transport equation for the distribution $f$ in the relaxation time approximation is given by (the power index $q$ on the l.h.s. of the following equation represents the entropic parameter),

\bea
\frac{\partial f^q}{\partial t} &=& - \frac{\left(f-f_{\text{eq}}\right)} {\tau} \nn\\
\frac{\partial f}{\partial t}    &=& - \frac{ \left(f^{2-q} - f_{\text{eq}} f^{1-q}\right)}   {q\tau},
\label{nebterta}
\eea
where $\tau$ is the relaxation time. 
Integrating Eq. \eqref{nebterta},
\bea
\int \frac{df}{\left(f^{2-q} - f_{\text{eq}} f^{1-q}\right)} &=&  \mathcal{K}-\theta  \nn\\
\frac{1}{q-1}\int \frac{dw} { \left(1- f_{\text{eq}} w^{-\frac{1}{q-1}} \right) } &=&  \mathcal{K}-\theta,\quad \nn\\
\text{where} \quad w\equiv f^{q-1}, \quad \theta = \frac{t}{q\tau}.
\eea
$\mathcal{K}$ is the integration constant that may be obtained from the boundary condition, $f(t=0)=f_{\text{in}}$, where $f_{\text{in}}$ is the
initial distribution. We expand the integrand in a negative binomial series and integrate.

\bea
\frac{1}{q-1} \int dw \left( 1 + f_{\text{eq}} w^{-\frac{1}{q-1}} + f_{\text{eq}}^2 w^{-\frac{2}{q-1}} +...\right) &=&  \mathcal{K}-\theta 
\quad \left( \left|f_{\text{eq}} w^{-\frac{1}{q-1}}\right| \equiv \left|\frac{f_{\text{eq}}}{f}\right| < 1 \right)
\nn\\ 
\Rightarrow \frac{f^{q-1}}{q-1} \mathlarger{\mathlarger{\sum}}_{s=0}^{\infty} \frac{(1)_s(1-q)_s}{s!(2-q)_s} \left(\frac{f_{\text{eq}}}{f}\right)^s
&=&  \mathcal{K} -\theta  \nn\\
\Rightarrow \frac{f^{q-1}}{q-1}   \, _2F_1\left(1,1-q;2-q;\frac{f_{\text{eq}}}{f}\right) &=&  \mathcal{K}-\theta, \nn\\
\label{nebtesol}
\eea
where $`(.)_s$' in the second line is the rising Pochhamer symbol given by, 
\bea
(a)_s = 
\begin{cases}
1 & s=0 \\
a(a+1).....(a+s-1) & \forall s>0,
\end{cases}
\eea
and $\, _2F_1$ is the hypergeometric function. The integration constant is given by,
\bea
\mathcal{K} = \frac{f_{\text{in}}^{q-1}}{q-1}   \, _2F_1\left(1,1-q;2-q;\frac{f_{\text{eq}}}{f_{\text{in}}}\right).
\eea
Hence, the solution of the nonextensive Boltzmann transport equation in the relaxation time approximation may be obtained once we solve
Eq.~\eqref{nebtesol} for $f$. This solution yields the modified (due to passage through the plasma) distribution. Although the solution of Eq.~\eqref{nebtesol} can be found using numerical methods, we can calculate 
approximate analytical solutions using the series expansion of the hypergeometric function given in the second line of 
Eq.~\eqref{nebtesol}. 
The zeroth order solution of Eq.~\eqref{nebtesol} ({\it i.e.} for $s=0$) can be found from the following equation,
\bea
\Psi_0 &=& f^{q-1}-(q-1)\left( \mathcal{K} -\theta \right)=0\nn\\
\Rightarrow f_0 &=& \left[(q-1)\left( \mathcal{K} -\theta \right)\right]^{\frac{1}{q-1}}.
\label{sol0}
\eea
A perturbative scheme to find higher order solutions has been developed in Ref.~\cite{TBPhysica}. However, in this work we consider
only the zeroth order solution. 



\subsection{$dN/dy$ ratio from NABTE}
The distribution $f_0$ is related to the Lorentz-invariant spectrum in the following way,
\bea
E \frac{d^3N}{d^3p} = \frac{gVE}{(2\pi)^3} f_0. 
\eea
Parameterizing $E$ and $p_z$ in terms of $y$ and $m_{\text{T}}$, the transverse momentum spectra can be obtained as,
\bea
\frac{d^2N}{dp_{\text{T}}dy} =\frac{gV}{(2\pi)^2} p_{\text{T}} ~m_{\text{T}}\cosh y ~f_0 \Rightarrow \frac{dN}{dy} = \int \frac{gV}{(2\pi)^2} p_{\text{T}} ~m_{\text{T}}\cosh y ~f_0 ~dp_{\text{T}}
\eea

By choosing an initial distribution and an equilibrium distribution that are, say, represented by the following functions,
\bea
f_{\text{in}} = \left(1+(q_{\text{in}}-1)\frac{E-\mu_{\text{in}}}{T_{\text{in}}} \right)^{-\frac{q_{\text{in}}}{q_{\text{in}}-1}}; 
~~~f_{\text{eq}} = \left(1+(q-1)\frac{E-\mu}{T} \right)^{-\frac{q}{q-1}}, 
\eea
the solution $f_0$ in Eq. \eqref{sol0} can be obtained. Hence, the $dN/dy$ ratio can be written as,
\bea
\frac{\int p_{\text{T}} ~m_{\text{T}}\cosh y ~f_{\text{0}} ~dp_{\text{T}}}
{\int p_{\text{T}} ~m_{\text{T}} \cosh y ~f_{\text{in}} ~dp_{\text{T}}}.
\label{rdndybte}
\eea
We compare the above ratio with experimentally observed $R_{dN/dy}$.

\section{Results and discussion}
\begin{figure}[!htb]
\minipage{0.42\textwidth}
\begin{center}
\hspace{-0.4in}
\includegraphics[width=\linewidth]{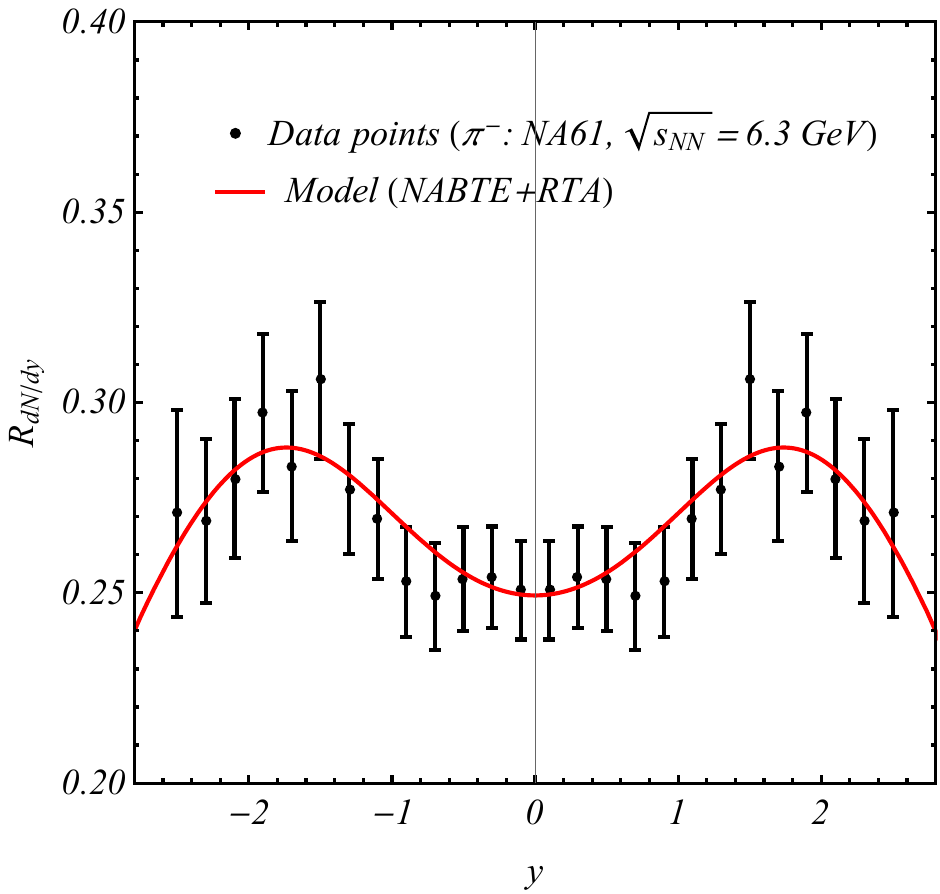}
\caption{Comparison of the ratio of $dN/dy$ in Pb-Pb and p-p collisions ($\sqrt{s}=6.3$ GeV) for the $\pi^{-}$ particles ($m$=0.139 GeV) with Eq.~\eqref{rdndybte} for the following parameter values: $q_{\text{in}}$=1.04, $q$=1.025, $T_{\text{in}}$=0.08 GeV, $T$= 0.051 GeV, $t/\tau$= 1.06, $\mu_{\text{in}}$= 0.12 GeV, $\mu$=0.092 GeV. We have put $N_{\text{coll}}$=808.}
\label{rdndybte6pt3}
\end{center}
\endminipage\hfill
\minipage{0.43\textwidth}
\vspace*{-0.0cm}
\hspace*{-2cm}
\includegraphics[width=\linewidth]{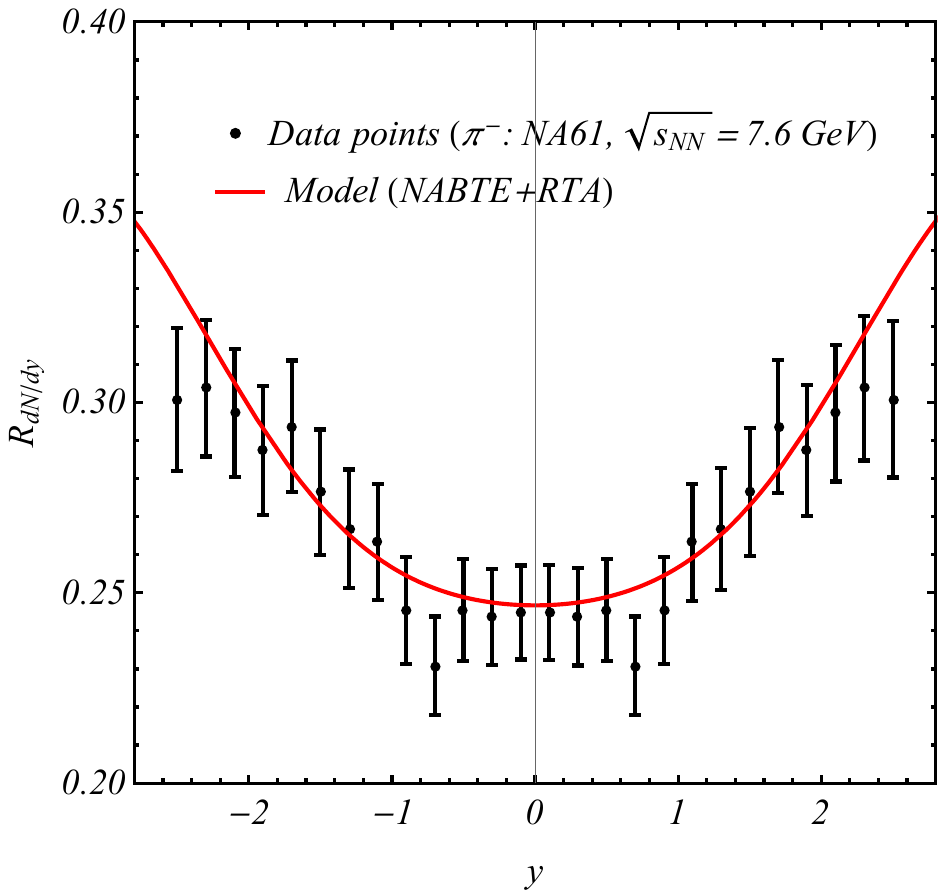}
\hspace{-1cm}
\caption{Comparison of the ratio of $dN/dy$ in Pb-Pb and p-p collisions ($\sqrt{s}=7.6$ GeV) for the $\pi^{-}$ particles ($m$=0.139 GeV) with Eq.~\eqref{rdndybte} for the following parameter values: $q_{\text{in}}$=1.012, $q$=1.009, $T_{\text{in}}$=0.11 GeV, $T$= 0.097 GeV, $t/\tau$= 0.824, $\mu_{\text{in}}$= 0.12 GeV, $\mu$=0.064 GeV.}
\label{rdndybte7pt6}
\endminipage\hfill
\end{figure}

\begin{figure}[!htb]
\minipage{0.42\textwidth}
\begin{center}
\hspace{-0.4in}
\includegraphics[width=\linewidth]{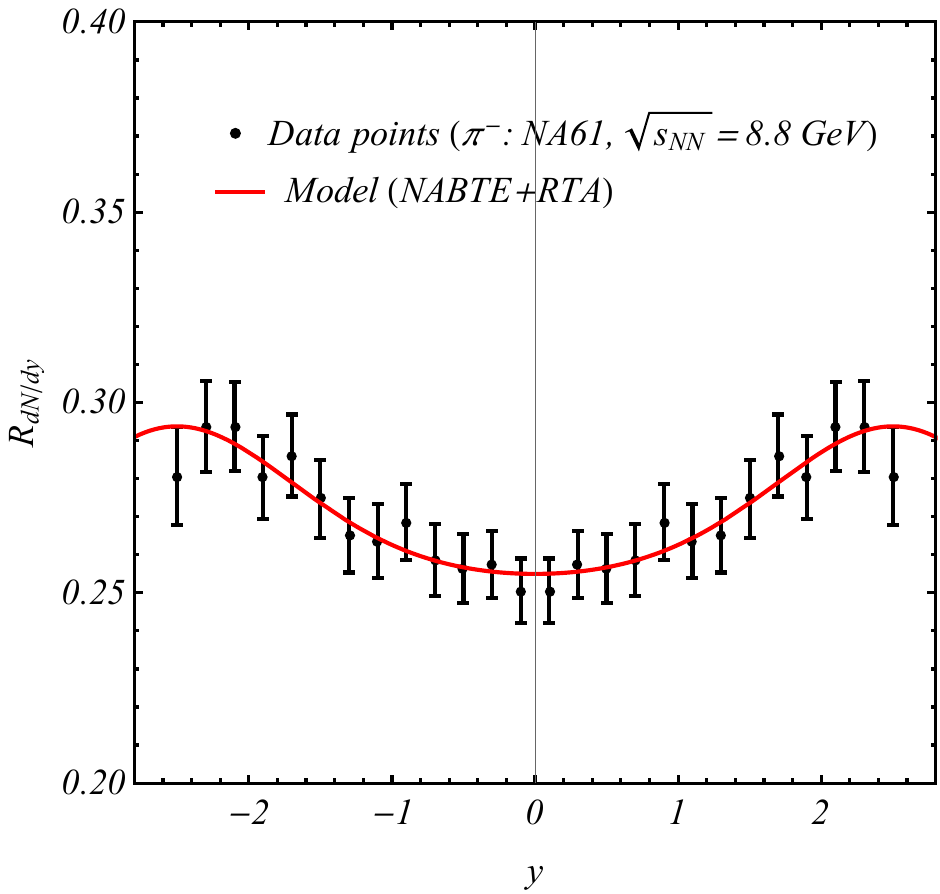}
\caption{Comparison of the ratio of $dN/dy$ in Pb-Pb and p-p collisions ($\sqrt{s}=8.8$ GeV) for the $\pi^{-}$ particles ($m$=0.139 GeV) with Eq.~\eqref{rdndybte} for the following parameter values: $q_{\text{in}}$=1.04, $q$=1.01, $T_{\text{in}}$=0.103 GeV, $T$= 0.08 GeV, $t/\tau$= 1.12, $\mu_{\text{in}}$= 0.2 GeV, $\mu$=0.1 GeV.}
\label{rdndybte8pt8}
\end{center}
\endminipage\hfill
\minipage{0.43\textwidth}
\vspace*{0.cm}
\hspace*{-2cm}
\includegraphics[width=\linewidth]{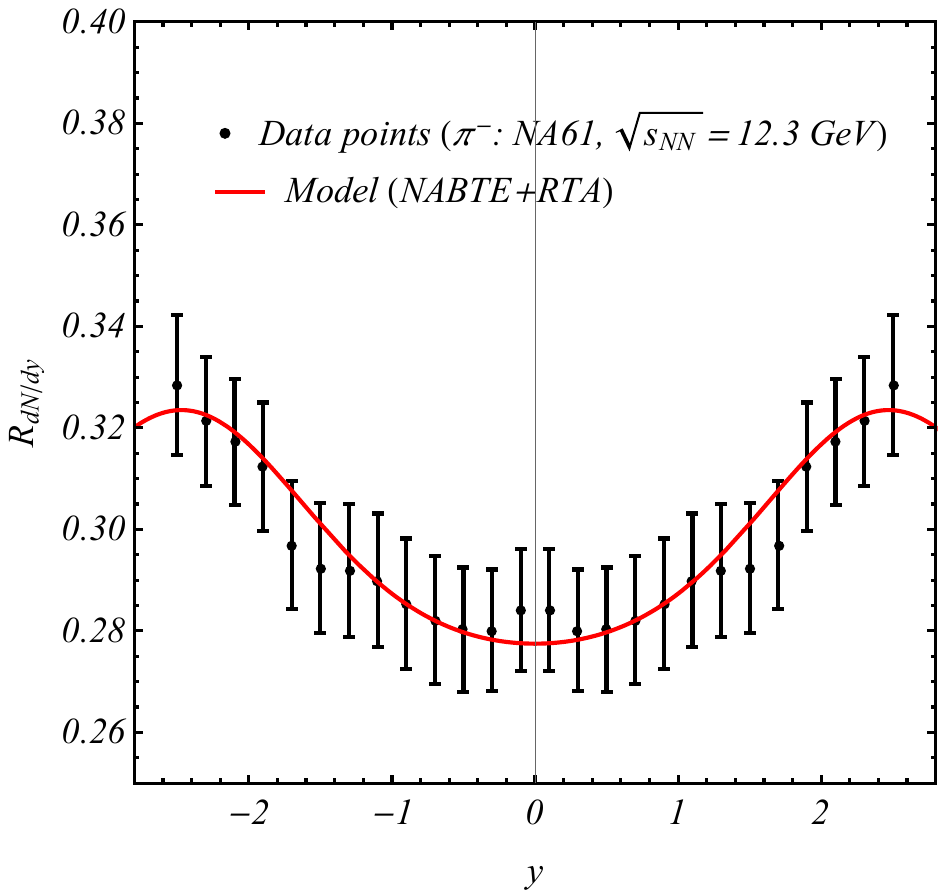}
\hspace{-1cm}
\caption{Comparison of the ratio of $dN/dy$ in Pb-Pb and p-p collisions ($\sqrt{s}=12.3$ GeV) for the $\pi^{-}$ particles ($m$=0.139 GeV) with Eq.~\eqref{rdndybte} for the following parameter values: $q_{\text{in}}$=1.05, $q$=1.01, $T_{\text{in}}$=0.108 GeV, $T$= 0.08 GeV, $t/\tau$= 1.04, $\mu_{\text{in}}$= 0.18 GeV, $\mu$=0.09 GeV.}
\label{rdndybte12pt3}
\endminipage\hfill
\end{figure}

\begin{figure}[htbp]
\begin{center}
\includegraphics[width=0.45\textwidth]{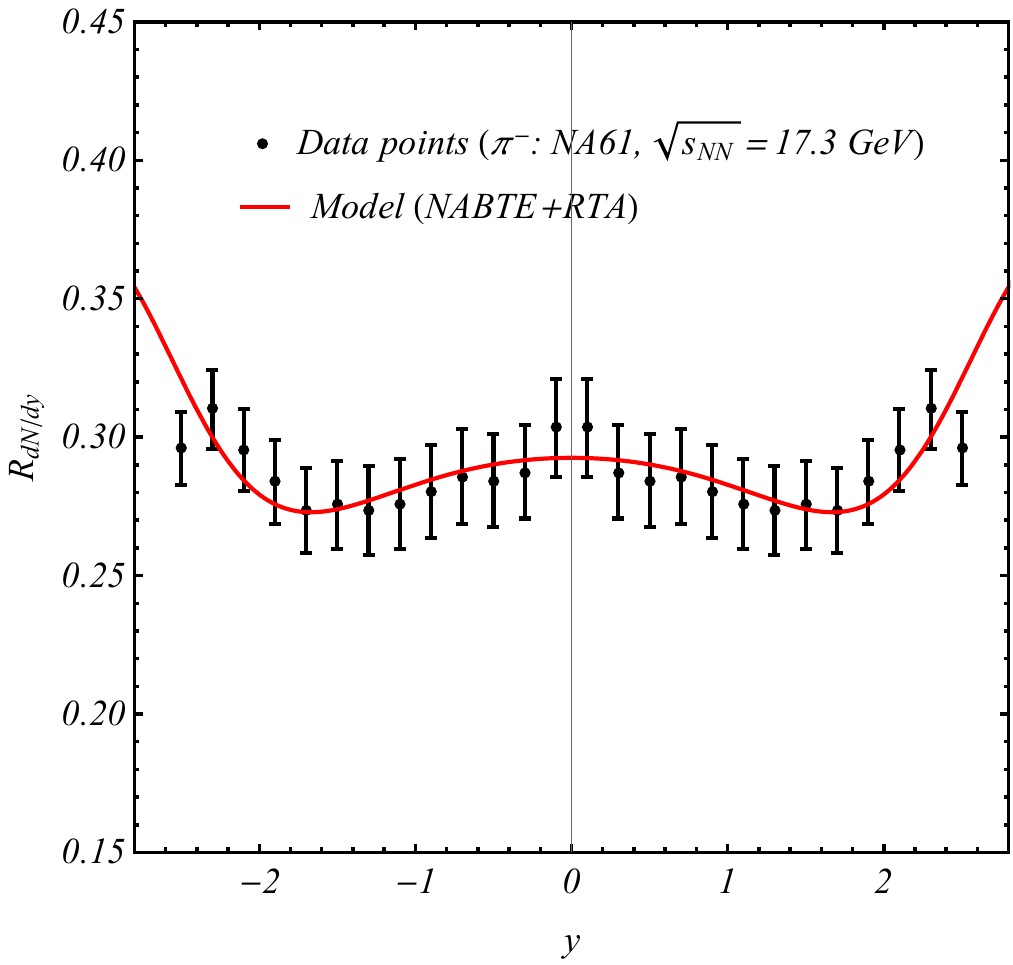}
\caption{Comparison of the ratio of $dN/dy$ in Pb-Pb and p-p collisions ($\sqrt{s}=17.3$ GeV) for the $\pi^{-}$ particles ($m$=0.139 GeV) with Eq.~\eqref{rdndybte} for the following parameter values: $q_{\text{in}}$=1.06, $q$=1.01, $T_{\text{in}}$=0.08 GeV, $T$= 0.09 GeV, $t/\tau$= 0.839, $\mu_{\text{in}}$= 0.16 GeV, $\mu$=0.04 GeV.}.
\label{rdndybte17pt3}
\end{center}
\end{figure}

In Figs.~\ref{rdndybte6pt3}-\ref{rdndybte17pt3} we compare our theoretical model (Eq.~\ref{rdndybte}) with experimental data \cite{NA61pp2017,NA49PbPb1,NA49PbPb2} for the negatively charged pions produced in proton-proton and central \footnote{Centrality is 0-7.2\% for $\sqrt{s_{NN}}$= 6.3-12.3 GeV and 0-5\% for $\sqrt{s_{NN}}$ = 17.3 GeV} Pb-Pb collisions at the center-of-mass energies $\sqrt{s_{NN}}$= 6.3, 7.6, 8.8, 12.3, and 17.3 GeV. 
In \textcolor{black} {these} figures we \textcolor{black}{show} the average number of binary collisions from GLISSANDO \cite{glissando} and $\langle N_{\text{coll}} \rangle $=808, 819, 825, 840, 900 respectively for the center-of-mass energies in the increasing order.
We observe that the model satisfactorily follow the experimental data points for the parameter values that are within the permissible range (e.g. $q<$ 4/3 for 3 momentum dimensions \textcolor{black}{\cite{Tsmbprd}}). \textcolor{black}{The parameter values are quoted in the captions of Figs. \ref{rdndybte6pt3}-\ref{rdndybte17pt3}. If we study these values,} we notice that the system starts evolving with an initial $q$ value and then relaxes to a lower but non-unity value of $q$. This signifies that the system of the pions relaxing inside the medium does not approach the Boltzmann-Gibbs limit. Consequently, their stationary distribution is still given by a power-law distribution. This observation is in keeping with a recent study that finds that the pions carry out L\'{e}vy walk inside the systems created in heavy-ion collisions \cite{CsanadNature}. L\'{e}vy walks are characterized by heavy-tailed random walks that indicate anomalous diffusion inside the system leading to a power-law stationary state \cite{wilkosada,TBplb}. This also justifies our choice of the nonadditive Boltzmann transport equation that has a power-law stationary state. We find that the values of the parameter $t/\tau$, a ratio of the freeze-out time to the relaxation time, vary around 1 indicating a possibility of the pions produced early in the collisions relaxing to a stationary state.
We also notice that the temperature value decreases except for the $\sqrt{s_{NN}}$=17.3 GeV plot in which there is a local maximum at around $y\approx0$. 

\section{Summary, conclusions, and outlook}
To summarize, in this work we have studied the longitudinal nuclear suppression factor, $R_{dN/dy}$. \textcolor{black} {First we carry out simulations as they are important to identify geometric baseline for our observable. We qualitatively expected the ratio to increase at higher forward/backward rapidities. Although simulations using FTFP$_{\text{BERT}}$ and HIJING yield such trend that is also seen in experimental data, EPOS results differ. We find this discrepancy among numerical simluations curious and worth investigating in the future. This discrepancy leads us to explore analytical models based on nonadditive statistics that describes particle transverse momentum spectra very well. We observe that for energies ranging between $\sqrt{s}$=6.3 GeV to 17.3 GeV, experimental data are explained by physically acceptable parameter values.} The first of the two analytical models we propose is based on the phenomenological Tsallis distribution that has been used to describe particle transverse momentum spectra. In this approach we derive a closed analytical formula for the rapidity spectra in terms of the hypergeometric function. As far as our knowledge goes, this analytical formula in Eq.~\eqref{dndyana} has been derived for the first time in the literature. Although the phenomenological approach may be successful in following the trend in experimental data, we have also proposed another transport-based model involving nonadditive Boltzmann transport equation in the relaxation time approximation. This equation yields a power-law stationary state that is the feature of a system where anomalous diffusion is prevalent. In this work we have considered the zeroth order solution obtained from a perturbative scheme proposed by one of the authors. However, it would be interesting to study how higher order solutions (provided in the appendix) influence parameter values. Although medium effects are incorporated through the relaxation time parameter, a more rigorous approach considering interaction matrix elements and a source term accounting for different other sources of pion production should be considered in a future work.

\section{Appendix: Higher order solutions}

The first order equation, whose solution we denote by $f_1(t)$, is given by,
\bea
\Psi_1 &=& f^{q-1}+ \left(\frac{1-q}{2-q} \right) f_{\text{eq}} f^{q-2} - (q-1)\left( \mathcal{K} -\theta \right) = 0.
\label{sol1}
\eea
Below we outline how an approximate analytical first order solution for the nonextensive Boltzmann transport equation
in the relaxation time approximation can be obtained following Ref. \cite{TBPhysica}. 
The solution of the first order equation can be written as a tiny increment over that of the zeroth order in the following way,
\bea
f_1=f_0+\epsilon_1, \quad |\epsilon_1| <<f_0.
\label{solf1eps}
\eea
Afterward, Eq.~\eqref{solf1eps} is put into Eq.~\eqref{sol1} that is expanded in terms of $\epsilon_1$ up to the first order (since $\epsilon_1$ is a small quantity). The resulting equation is solved for $\epsilon_1$ and one gets $f_1$ in terms of $f_0$ whose analytical form is already known from Eq.~\eqref{sol0}. This gives us the following expression for the solution of the
first order equation,
\bea
f_1 \approx f_0 + \frac{f_0}{f_0+f_{\text{eq}}} \left[ \frac{f_{\text{eq}}}{2-q} + \frac{f_0}{1-q} + f_0^{2-q} \left( \mathcal{K} -\theta \right) \right]. \nn\\
\label{nebtertaapproxsol}
\eea
Following Eq.~\eqref{solf1eps}, the first order and the higher order solutions can be represented by the following recursion,
\bea
~~~~~~~~~~~~~~~~~~~~~~~~~
f_i = f_{i-1} +\epsilon_i, \quad i=1,~2,~3,...,
\eea
where $\epsilon_i^{\text{s}}$ are calculated from the following equation,
\bea
\epsilon_i = \frac{ f_{i-1}  }
{  \mathlarger{\sum}_{r=0}^i  f_{\text{eq}}^r f_{i-1}^{i-r} }
\left( f_{i-1}^{i+1-q} (\mathcal{K}-\theta) + \mathlarger{\mathlarger{\sum}}_{r=0}^i \frac{f_{\text{eq}}^r f_{i-1}^{i-r}} {r+1-q} \right). \nn\\
\eea

\section*{Acknowledgements}
TB acknowledges funding from the European Union's HORIZON EUROPE programme, via the ERA Fellowship
Grant Agreement number 101130816. During preparation of this publication we used the resources of the Centre 
for Computation and Computer Modelling of the Faculty of Exact and
Natural Sciences of the Jan Kochanowski University in Kielce, modernised
from the funds of the Polish Ministry of Science and Higher Education in
the ``Regional Excellence Initiative'' programme under the project
RID/SP/00015/2024/01.


\end{document}